\documentclass[prl,aps,showpacs,twocolumn,draft]{revtex4}
\usepackage{epsf}
\begin{document}
\bibliographystyle{prsty}
\title{Antiferromagnetism and hole pair checkerboard in the vortex state of high Tc superconductors}
\author{Han-Dong Chen${}^{1}$, Jiang-Ping Hu${}^{1}$, Sylvain Capponi${}^{1,2}$, Enrico Arrigoni${}^{1,3}$ and Shou-Cheng Zhang${}^{1}$}

\address{\it ${}^{1}$Department of Physics, McCullough Building,
Stanford
  University, Stanford  CA~~94305-4045\\
   ${}^{2}$Universit\'e Paul Sabatier, Laboratoire de
Physique Quantique, 118 route de Narbonne, 31062 Toulouse,
France\\
${}^{3}$Institut f\"ur Theoretische Physik, Universit\"at
W\"urzburg, Am Hubland, 97074 W\"urzburg, Germany}

\begin{abstract}
We propose a microscopic state for the vortex phase of BSCO
superconductors. Around the vortex core or above $H_{c2}$, the
$d$ wave hole pairs form a checkerboard localized in the
antiferromagnetic background. We discuss this theory in
connection with recent STM experiments.
\end{abstract}
 \maketitle

Since the theoretical prediction of the antiferromagnetic (AF) order in
the vortex state of the high $T_c$ superconductors (SC)\cite{zhang1997,arovas1997},
tremendous
experimental\cite{lake2001,lake2002,miller2001,mitrovic2001,hoffman2002,khaykovich2002,mitrovic2002}
and theoretical\cite{bruus1998,ogata1999,hu1999,demler2001,hu2001,lee2001,chen2001,sachdev2002,zhu2002,polkovnikov2002}
progress has been made in recent years.
Neutron scattering experiments\cite{katano2000,lake2001,lake2002,khaykovich2002} in the
$La_{2-x}Sr_xCuO_4$ compound first reported enhanced magnetic scattering induced by the magnetic
field in the vortex state. The field induced moment scales linearly (with possible
logarithmic corrections) with the
external field, or the number of vortices in the system, in accordance with
theoretical predictions\cite{arovas1997,demler2001}.
The $\mu$SR experiment\cite{miller2001}
in the underdoped $YBa_2Cu_3O_{7+\delta}$ compound reveals static AF order in the vortex core. The NMR experiment.\cite{mitrovic2001,mitrovic2002} in the $YBCO$ compound under high magnetic field shows enhanced AF ordering, predominantly from the region inside the vortex core. These experiments have important implications on our theoretical understanding of high $T_c$ superconductivity. In the $SO(5)$ theory, the transition from the AF Mott insulator phase to the SC phase is understood in terms of the rotation of the $SO(5)$ superspin vector as the chemical potential or doping level is varied\cite{zhang1997}. Recent experiments in the AF vortex state show that the $SO(5)$ rotation picture can be validated in real space, as a function of the distance from the center of the vortex core, where the superspin rotates continuously from AF to SC directions.

While the salient features of the recent AF vortex experiments are in
overall agreement with the theoretical predictions, there are additional experimental features worth exploring. A striking feature is revealed in the STM experiments by Hoffman {\it et al.}\cite{hoffman2002}, where the local density of states (DOS) near the vortex core show a two dimensional checkerboard-like modulation with a $4a\times 4a$ charge unit cell. Here $a$ is the dimension of the $CuO_2$ unit cell.  This modulation decays {\it exponentially} away from the center of the vortex core, with a decay length of about $10a$. More recently, a similar pattern has also been seen
in the absence of the applied magnetic field\cite{howald2002}, possibly induced by the impurities at the surface.
Understanding the four unit cell checkerboard modulation structure around the vortex core is the main focus of our present work.

One possible explanation of the STM experiments is the formation of the stripe AF order\cite{zaanen2000,emery1999} around the vortex core or impurities.
If the stripes are {\it static}, the question is how the 1D stripes on two different layers of the bi-layer $Bi_2Sr_2CaCu_2O_{8+\delta}$ material are oriented relative to each other. If we assume a strong coupling between the bi-layers, one would naturally expect the stripes on two different layers to be oriented along the same direction. This state breaks the rotational symmetry and is inconsistent with the 2D checkerboard pattern observed in the STM experiments.
On the other hand, it is possible that the stripes on two different layers are oriented along the orthogonal directions. In this case, the STM tunneling should still predominantly probe the upper layer, and observe a strongly anisotropic quasi-1D pattern. Therefore, the {\it static} 1D stripe picture can be experimentally falsified as the instrumental resolution improves. Current observations could be consistent with a picture of dynamically fluctuation stripes. The question is the
time scale of the fluctuation. As the magnetic field increases, these fluctuations are expected to slow down, and eventually, a static 1D ordering pattern should emerge. This scenario could be tested in the underdoped materials or in the $Nd_{2-x}Ce_xCuO_{4-\delta}$ materials, where $H_{c2}$ is within experimental reach.
Besides the stripe state, one could also try to construct a Wigner crystal state of the doped holes in a AF background. Such an insulating state can only be realized for rational doping fractions. The optimally doped system is close to $x=1/8$ doping, with one hole per eight sites. In this case, the holes can form a $\sqrt{8} a \times \sqrt{8} a$ superlattice uniformly distributed over the AF spin background. Such a state has the full four-fold rotational symmetry as observed in the STM experiment. Unfortunately, the charge unit cell in this case is smaller than the experimentally observed $4a \times 4a$ structure. Finally, the simplest explanation of the
checkerboard charge structure around the impurities could be the Friedel oscillation associated with the fermi surface nesting\cite{lee2002}. The observed $4a \times 4a$ charge pattern could be produced by the quasi-particles at the anti-node.
However, these quasi-particles are gapped, and it is not clear if
they can explain the low energy feature at $7meV$ observed in the
vortex core experiment.

\begin{figure}[h]
\centerline{\epsfysize=6cm \epsfbox{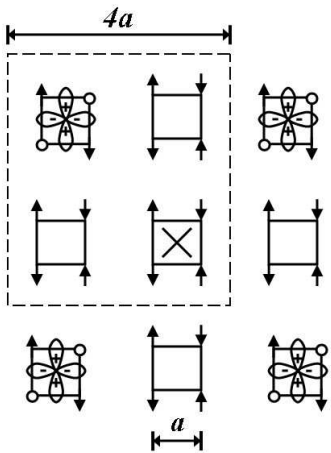}} \caption{
Illustration of the $d$ wave hole pair checkerboard state at
$x=1/8$. In this state, the $d$ wave hole pairs occupy every four
non-overlapping plaquettes on the original lattice. The charge
unit cell is $4a\times 4a$. The $SO(5)$ model is defined on the
center of the non-overlapping plaquettes. Such a state could be
realized around the vortex core, whose center is depicted by the
cross. In the actual realization of this state, the hole pair can
be much more extended, and the AF ordering could be much reduced
from the classical value.
 }
 \label{fig1}
\end{figure}

In this work, we propose a specific microscopic state around the
vortex core and above $H_{c2}$ for the $YBCO$ and $BSCO$ class of
superconductors. The insulating version of this state at doping
$x=1/8$ is depicted in Fig. \ref{fig1}. In this state, the $d$
wave hole pairs form a Wigner crystal, dubbed a $d$ wave hole pair
checkerboard. The center-of-mass of the $d$ wave hole pair is
located at the center of every four plaquette on the original
lattice. This state has the $4a\times 4a$ checkerboard symmetry as
observed in the experiment, and the doping level for the
insulating $x=1/8$ state is reasonably close to the optimal doping
level of the cuprates. Therefore, by forming the Wigner crystal
state of the hole pairs rather than the holes themselves, the
doping level can be compatible with the observed size of the
charge unit cell. Here the $d$ wave pair is drawn on the plaquette
for illustrative purposes. In the actual system, the pair size can
be more extended. In this Letter, we discuss the case where
the AF regions are connected, therefore, antiphase domain walls would be
energetically unfavorable. However, the antiphase domain
walls would be possible if the hole pairs are more extended along the
four-fold symmetry direction, and if there is an effective
coupling among the triplet magnons on next-nearest-neighbor plaquettes.
We propose that the hole pair checkerboard state is energetically competitive
in the microscopic $t-J$ model with short ranged Coulomb interaction.
This state takes advantage of the underlying AF exchange energy in
the background, while the holes form singlet $d$ wave bonds to
minimize their exchange energy as well. Finally, the kinetic
energy of the hole pair competes with the Coulomb interaction. At
$x=1/8$, if the Coulomb interaction is stronger than the kinetic
energy of the hole pair, the hole pair checkerboard state could
win. We believe that the $YBCO$ and $BSCO$ materials are in the
regime where the kinetic energy of the hole pairs favor the
uniform SC state. However, when the applied magnetic field impedes
the coherent hole pair motion either around the vortex core or in
the state above $H_{c2}$, we argue that the hole pair checkerboard
state can be realized, either in the static or dynamically
fluctuating form.
Therefore, the central idea behind this work is
that the magnetic field destroys the phase coherence of the hole
pair by localizing them into a crystal, without breaking the pair.

The hole pair checkerboard state arises naturally from the
projected $SO(5)$ model on a
lattice\cite{zhang1999,altman2001,dorneich2002}. In this approach,
the original lattice is divided into non-overlapping plaquettes,
depicted by solid squares in Fig. \ref{fig1}. In the spirit of
real-space renormalization, we retain only five low energy states
on the plaquette, the half-filled singlet state, denoted by
$|\Omega\rangle$, the triplet magnon state
$t^\dagger_\alpha|\Omega\rangle$ and $d$ wave hole pair state
$t^\dagger_h|\Omega\rangle$ with 2 holes on the plaquette. The $d$
wave hole pair on a plaquette can be justified from the
microscopic models\cite{scalapino1996}, but a more extended hole
pair can also be incorporated into our model. We define the center
of the non-overlapping plaquettes to be the lattice sites of the
projected $SO(5)$ model, which has the spacing of $2a$. Here we
follow the notations and the conventions of reference
\cite{zhang1999}. The projected $SO(5)$ model is defined in terms
of the following Hamiltonian.
\begin{eqnarray}
H &=&
\Delta_s \sum_{ x } t_\alpha^\dagger t_\alpha(x) +
\tilde\Delta_c \sum_{ x } t_h^\dagger t_h(x) \nonumber \\
&-& \frac{J_s}{2} \sum_{\langle xx'\rangle} (t^\dagger_\alpha+t_\alpha)(x)
(t^\dagger_\alpha+t_\alpha)(x') \nonumber \\
&-& \frac{J_c}{2} \sum_{\langle xx'\rangle} e^{i\phi_{ij}} t^\dagger_h(x) t_h(x')+h.c. \nonumber \\
&+& V_1 \sum_{\langle xx'\rangle} \rho(x) \rho(x') + V_2
\sum_{\langle\langle xx'\rangle\rangle} \rho(x) \rho(x'),
\label{H}
\end{eqnarray}
where $\rho(x)=t_h^\dagger t_h(x)$ is defined as the hole-pair density.

 As described in reference \cite{zhang1999}, the $\Delta_s$
term describes the effective spin gap energy for the magnons, the
$\tilde\Delta_c$ term describe the effective chemical potential
for the hole pairs, while the $J_s$ and $J_c$ terms describe the
effective AF and SC Josephson couplings respectively. Here
$e^{i\phi_{ij}}$ corresponds to the phase introduced by the
magnetic field in the vortex state, and we have also included the
nearest neighbor and next nearest neighbor Coulomb interactions
$V_1$ and $V_2$. This model can be approximately derived from the
microscopic Hubbard and $t-J$ model by a renormalization group
transformation\cite{altman2001}. The model without the Coulomb
terms have been studied extensively by numerical calculations, and
the phase diagram has been determined\cite{dorneich2002}.

If one neglects the spin degree of freedom, the $SO(5)$ model
reduces to the hard core boson model on a two dimensional square
lattice. Even this simple model has a rich phase diagram as the
density, the kinetic and the potential energies are
varied\cite{bruder1993,hebert2002}. Besides the superfluid ground
state, the system can also develop charge-density-wave order
and supersolid order in the ground state. In the strong coupling
limit at half-filling, the ground state develops checker-board
charge order where every other site is occupied by a boson. At
quarter-filling, a similar checker-board pattern develops where
every four sites is occupied by a boson. This state is stabilized
by the nearest neighbor and the next-nearest neighbor interactions
$V_1$ and $V_2$. Including the AF spin order and translating the
hole density and lattice size from the $SO(5)$ lattice model to
the original lattice, we find that the $1/4$ filled checkerboard
charge ordering of the hard-core bosons corresponds exactly to the
$d$ wave pair checkerboard state illustrated in Fig. \ref{fig1}.
However, we shall see that the AF exchange energies also play an
important role in determining the competition between the various
states.

We consider the following mean field state for the Hamiltonian (\ref{H}),
\begin{equation}
|\Psi\rangle = \prod_x (\cos\theta+\sin\theta(\cos\alpha
t^\dagger_\alpha +\sin\alpha e^{i\phi} t^\dagger_h)) (x)
|\Omega\rangle \label{variational}
\end{equation}
and minimize the energy $E=\langle\Psi|H|\Psi\rangle$ as a
function of the variational parameters $\theta(x),\alpha(x)$ and
$\phi(x)$. At quarter filling of the hole pair bosons, (which
corresponds to $x=1/8$ in the actual system), the mean field
theory predicts a quantum phase transition at $V_1+V_2=V_c\equiv
6J$ for $J_c=2J_s\equiv 2J$. For $V_1+V_2<V_c$, the system is in
an uniform SC state. On the other hand, for $V_1+V_2>V_c$, a hole
pair checkerboard state depicted in Fig. \ref{fig1} is favored. We
shall choose our parameters such that the system is in an uniform
SC state, but it is close to the hole pair checkerboard state.

In order to reduce the number of variables, we have assumed
four-fold rotational symmetry as well as symmetry by reflection
along the diagonals. We can further simplify the problem by fixing
the SC phase $\phi$ assuming that the magnetic field is uniform in
a region of radius equal to the London penetration depth (here, we
present data with $30a$ but we have checked that this effect is
very small).

The boundary sites are taken to be in a bulk SC phase
with a fixed hole-pair density, which is taken to be the same as
the averaged one. We
fixed the center of the vortex core to be fully anti-ferromagnetic, {\it i.e.},
$\alpha=0$ and $\theta=\pi/4$.

\begin{figure}[h]
\centerline{\epsfysize=3.8cm \epsfbox{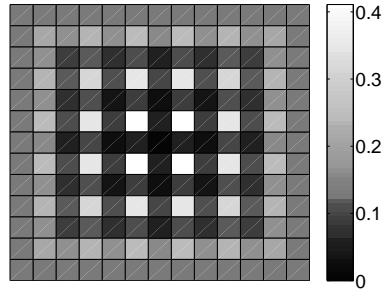} } \caption{ Charge
distribution around the vortex core. Parameters of the Hamiltonian
are: $V_1/J=3$, $V_2/J=2.2$, $\Delta_S/J=1.2$ and
$\tilde\Delta_c/J=-0.12$. The center of the figure is the center
of the vortex. The grid size is $2a$.}
 \label{fig2}
\end{figure}

In Fig.\ref{fig2}, we present the result of charge distribution
which shows a clear checkerboard pattern. The contrast decreases
very quickly away from the vortex core over a distance $\sim 10 a$
which is roughly consistent with the experimental
value~\cite{hoffman2002}. We would like to emphasize that this
checkerboard pattern persists for smaller interactions $V_1$ and
$V_2$, but with a smaller contrast. For sufficiently small
contrast, the modulation of the charge density may be hard to
detect in the integrated LDOS\cite{hoffman2002}.

\begin{figure}[h]
\centerline{\epsfysize=3.3cm \epsfbox{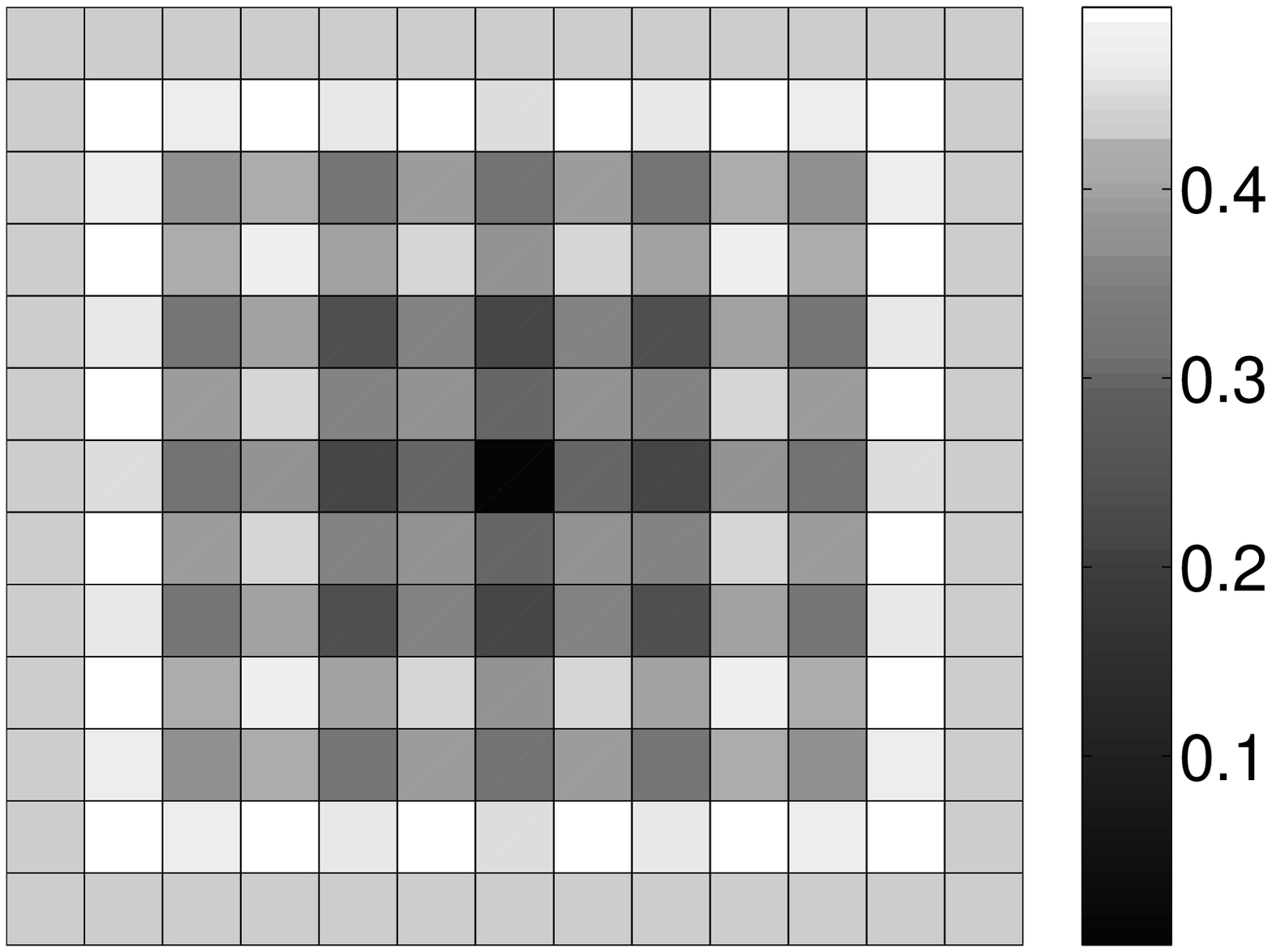} }
\centerline{\epsfysize=3.3cm \epsfbox{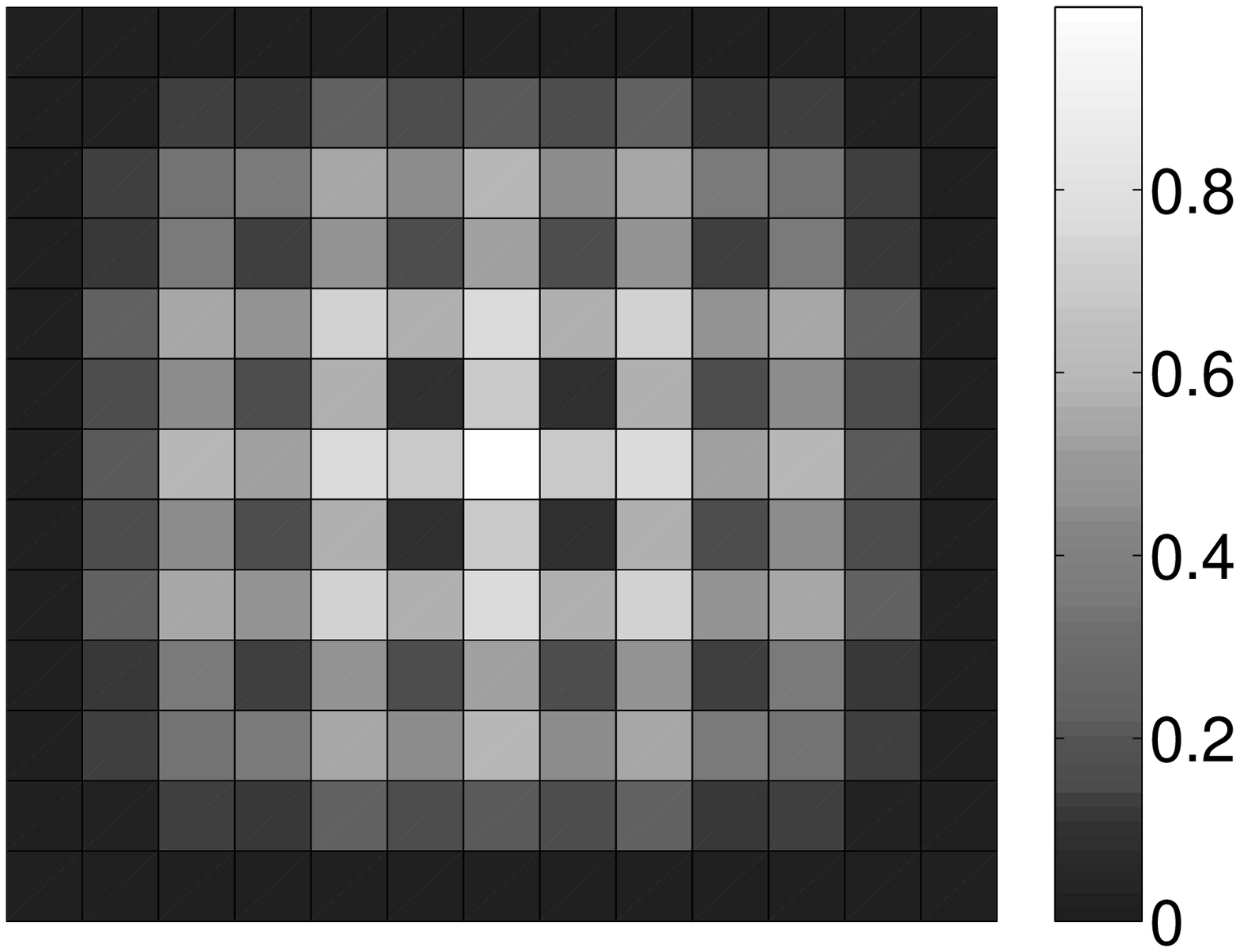} }
\centerline{\epsfysize=3.3cm \epsfbox{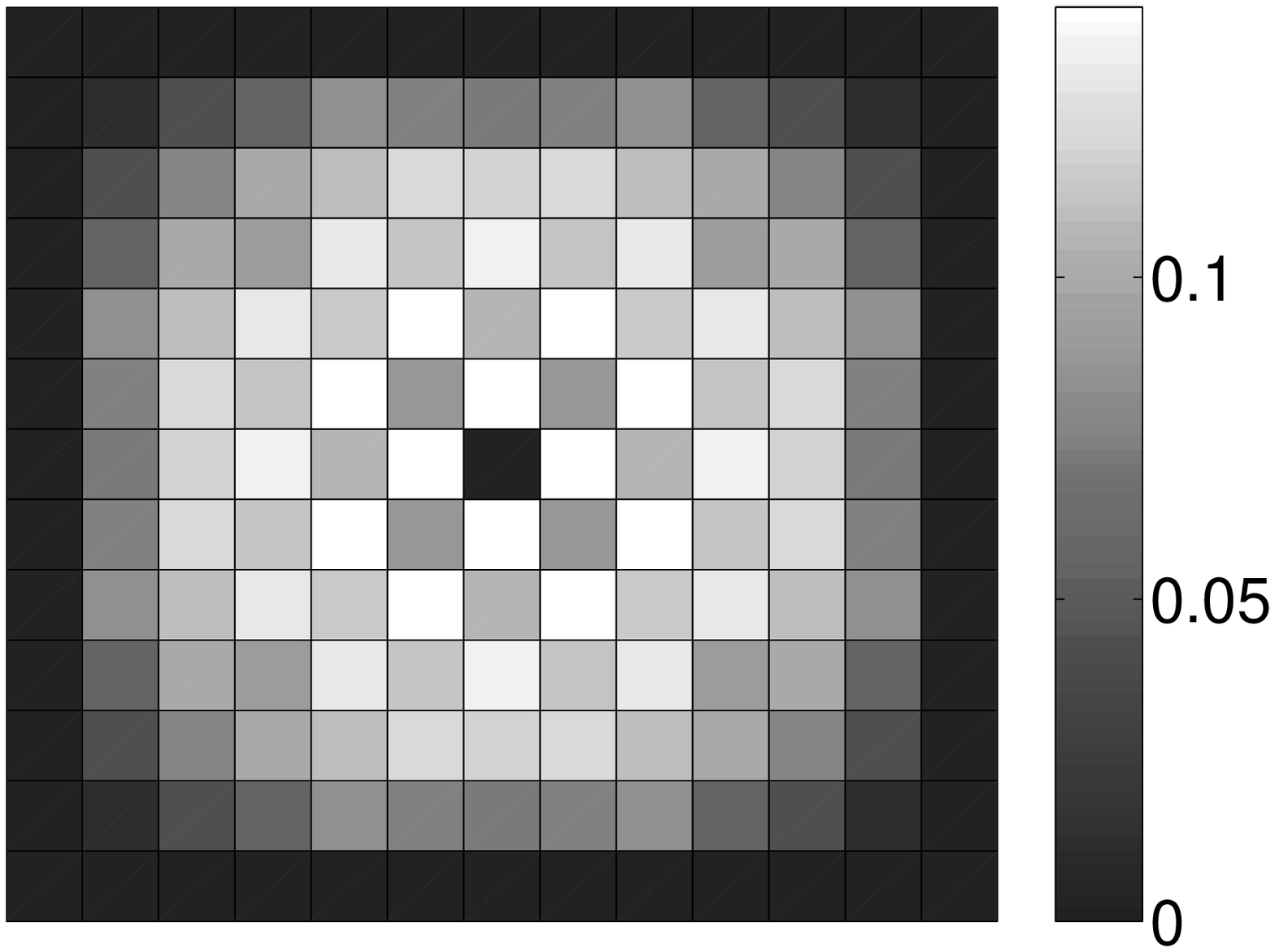} } \caption{From top
to bottom: Distribution of the SC, AF and $\pi$ order parameters
around the vortex core. The relative size of the AF and SC regions
can be varied by varying $J_s/J_c$. Notice that the $\pi$ order
parameter is only non-zero in the region where the AF and SC order
parameters overlap.
 }
 \label{fig3}
\end{figure}

We also present the profiles of the AF order parameter $\langle
n_2\rangle$, the SC order parameter $\langle n_1\rangle$ and the
$\pi$ order parameter $\langle
iL_{25}\rangle/2$\cite{zhang1997} in Fig.\ref{fig3}.
Notice that the $\pi$ order parameter is only non-zero in the
region where the AF and SC order parameters overlap. In our
projected SO(5) model, the $SO(5)$ orthogonality
condition\cite{zhang1997,hu2001}
\begin{equation}
\langle L_{ab}\rangle\langle n_c\rangle +\langle
L_{bc}\rangle\langle n_a\rangle +\langle L_{ca}\rangle\langle
n_b\rangle=0
\end{equation}
is satisfied at every site. This relation predicts a quantitative
relation among these three order parameters and the hole density,
$\rho\langle n_2\rangle + \langle L_{25}\rangle\langle n_1\rangle
= 0 $,which is satisfied by our numerical results of the order
parameters. The relative size of the AF and SC regions is
determined by the ratio of $J_s/J_c$, and can be varied to account
for the experimental observations at low temperature.

We shall now discuss the experimental consequences of our theoretical proposal. Our proposal can be tested most directly in the vortex state, or above $H_{c2}$, by measuring the static charge structure factor corresponding to a $4a\times 4a$ unit cell with $XY$ rotational symmetry. The magnetic structure factor could either be a commensurate $(\pi,\pi)$ peak with an amplitude modulation of four unit cells or a shifted peak corresponding to an anti-phase modulation.
Since the energy difference between these two types of magnetic order is very small, we conjecture that both types could occur in difference compounds of the high $T_c$ cuprates. Our proposal differs in detail from the stripe picture from the fact that the ordering pattern has square symmetry and from the fact that the ordering charge is the hole pair rather than single holes. Our proposed state is also consistent with the $\mu$sR and the NMR experiments in the vortex state of the $YBCO$ system\cite{miller2001,mitrovic2002}. The $\mu$sR experiment reports commensurate $(\pi,\pi)$ AF order inside the vortex core\cite{miller2001}. Because of the modulation of the AF order parameter amplitude, oxygen sites, for examples those located in between the singlet hole pair and the AF plaquettes, can still detect a spin signal\cite{miller2001,mitrovic2002}. 
In this work we primarily focused on the charge ordering in the vortex state. Similar arguments may be applied to the impurity states as well\cite{howald2002}. However, we caution here that if the impurity is pair
breaking, the idea of a order state of hole pairs may not be directly applicable. We would also like to call attention to ref. {\cite{kim2001}, where experimental evidence for Wigner lattice was reported.

We also encourage the {\it detailed study } of this hole pair checkerboard state in the numerical calculations of the $t-J$ model, including effects of the Coulomb interaction\cite{Dagotto1995}.  While we do not propose this state to be realized in the ground state for physical range of parameters in the absence of the magnetic field, it is interesting to test if this
state is realized in the vortex state or above $H_{c2}$. It is also instructive to determine parameters for which this state is realized in the ground state, and see how far these parameters are from the physically reasonable values. This way, we can estimate the perturbations required for this state to reveal itself.

 We would like to acknowledge useful discussions with Drs. J.
Berlinsky, S. Davis, E. Dagotto, E. Demler, W. Hanke, C. Kallin,
A. Kapitulnik, S. Kivelson and D.H. Lee. This work is supported by
the NSF under grant numbers DMR-9814289. HDC and JPH are also
supported by a SGF, and E.A. by a Heisenberg fund of the DFG (AR 324/3-1) and by KONWIHR CUHE.


\end{document}